\documentclass{pos}
\usepackage{epsfig,graphicx}
\usepackage{amssymb}

\newcommand{\beq}{\begin{equation}}
\newcommand{\eeq}{\end{equation}}
\newcommand{\be}{\begin{equation}}
\newcommand{\ee}{\end{equation}}
\newcommand{\ba}{\begin{array}}
\newcommand{\ea}{\end{array}}
\newcommand{\beqa}{\begin{eqnarray}}
\newcommand{\eeqa}{\end{eqnarray}}
\newcommand{\bea}{\begin{eqnarray}}
\newcommand{\eea}{\end{eqnarray}}
\newcommand{\lsim}{\stackrel{<}{_\sim}}
\newcommand{\gsim}{\stackrel{>}{_\sim}}

\newcommand{\cA}{{\cal A}}
\newcommand{\cO}{{\cal O}}

\newcommand{\cL}{{\cal L}}

\newcommand{\no}{\nonumber}
\newcommand{\yuk}{{\lambda}}
\newcommand{\LLN}{{\Lambda_{\rm LN}}}
\newcommand{\LLFV}{{\Lambda}}

\newcommand{\sla}{\! \! \! \! \!  /~}

\def\npb#1#2#3{    {\it Nucl. Phys. }{\bf B #1} (#2) #3}
\def\plb#1#2#3{    {\it Phys. Lett. }{\bf B #1} (#2) #3}

\def\prd#1#2#3{    {\it Phys. Rev. }{\bf D #1} (#2) #3}

\def\epjc#1#2#3{   {\it Eur. Phys. J. }{\bf C #1} (#2) #3}

\title{Effective Theories for Flavour Physics beyond 
the Standard Model}

\ShortTitle{EFT for Flavour Physics BSM}

\author{\speaker{Gino Isidori}\thanks{Work supported by the EU under contract 
MTRN-CT-2006-035482.}\\
        INFN, Laboratori Nazionali di Frascati, Via E. Fermi 40
I-00044 Frascati, Italy.\\
        E-mail: \email{Gino.Isidori@lnf.infn.it}}

\abstract{We discuss the role of flavour physics in building 
effective theories at the TeV scale. Particular attention is devoted 
to the Minimal Flavour Violation hypothesis, both in the quark and 
in the lepton sector. Alternative flavour-protection mechanisms, 
such as the hierarchical fermion profiles of
models with a warped fifth dimension, 
are also briefly discussed.
}

\FullConference{International Workshop on Effective Field Theories: 
 from the pion to the upsilon \\		
February 2-6 2009\\	Valencia, Spain}

\begin{document}

\section{Introduction: the SM as an Effective Field Theory}
Most of this conference has been devoted to discuss 
Effective Field Theories (EFT) describing the low-energy limit of the 
Standard Model (SM); however, the SM
itself should be considered 
as the low-energy limit of a more fundamental theory.
In this perspective, three key issues need  to be addressed: 
(1) which is the cut-off, (2) which are the 
light degrees of freedom, and (3) which are the symmetries 
of the SM viewed as an EFT. 

The first two questions are intimately related to the breaking 
to the $SU(2)_L \times U(1)_Y$ electroweak gauge symmetry. 
Experiments provide a clear indication of a spontaneous breaking of 
$SU(2)_L \times U(1)_Y$,  characterised by the 
order parameter $v=(\sqrt{2}G_{F})^{-1/2}\simeq246$~GeV. 
If we do not include a Higgs field among the light degrees 
of freedom of the theory, assuming that the electroweak symmetry
breaking occurs because of some new strong dynamics at high energies, 
the cut-off of the EFT cannot be larger than $4\pi v \approx 3$~TeV,
in close analogy to what happens in  Chiral Perturbation Theory. 
If we do assume that there is a Higgs field, with renormalisable 
potential and non-trivial electroweak vacuum, the EFT becomes 
renormalisable and it is not obvious how to determine its cut-off.
However, also in this case it is {\em natural} to assume 
that the EFT breaks down around or below the TeV, 
because of the instability of the Higgs mass term under 
quantum corrections. 

What I will try to address in this talk is the last of the three
questions listed above: which are the symmetries, and in particular 
which is the flavour symmetry and the related symmetry-breaking pattern,
of the SM viewed as an effective theory with a few TeV cut-off. 
The assumption that the cut-off of the 
theory is around a few TeV is consistent with the measurements 
of flavour-conserving electroweak precision 
observables (EWPO). More precisely, the bounds on gauge-invariant 
higher-dimensional operators constructed in terms of 
light degrees of freedom (with or without the Higgs boson)
are consistent with an effective cut-off in the 
TeV range.\footnote{~In the Higgsless case this statement 
is true only if we impose the so-called $SU(2)$ {\em custodial symmetry} 
on the leading operators responsible for the electroweak symmetry breaking.} 
The situation is very different if we look at operators 
contributing to flavour-changing observables. Here the good 
agreement between SM expectations and data leads to 
bounds on the EFT cut-off which are orders of magnitude 
higher, unless some protective flavour symmetry is invoked 
to suppress the corresponding couplings. 

The need of additional global symmetries does not come 
only from the flavour sector: the most stringent 
constrains on the EFT arise by barion- and
lepton-number violating processes. However, since $B$ and $L$ are 
exact symmetries of the SM Lagrangian, in this case the 
problem can easily be solved promoting $B$ and $L$ 
to be exact symmetries of the new dynamics 
at the TeV scale. The peculiar aspects of flavour 
physics is that there is no exact flavour symmetry 
in the low-energy theory.  In this case it is 
not sufficient to invoke a flavour symmetry 
for the EFT. We also need to specify how this symmetry 
is broken in order to describe the observed low-energy 
spectrum and, at the same time, be in agreement 
with the precise experimental tests
of flavour-changing processes.

\section{The flavour problem}
\label{sect:flav_prob}

Assuming for simplicity that there is a single elementary 
Higgs field, responsible for the $SU(2)_L \times U(1)_Y \to U(1)_Q$
spontaneous breaking, the Lagrangian of our EFT can be written as follows
\be
\cL_{\rm eff} = \cL^{\rm SM}_{\rm gauge} + \cL^{\rm SM}_{\rm Higgs} + 
\cL^{\rm SM}_{\rm Yukawa}
+\Delta \cL_{d > 4}~,
\ee
where $\Delta \cL_{d > 4}$ denotes the series of higher-dimensional 
operators constructed in terms of SM fields and invariant under 
the SM gauge group:
\be
\Delta \cL_{d > 4}  = ~\sum_{d > 4} ~\sum_{n=1}^{N_d} ~
\frac{c^{(d)}_n}{\Lambda^{d-4}} \cO^{(d)}_n 
\ee
As discussed in the introduction, we should 
expect $\Lambda=O({\rm few~TeV})$ and $c^{(d)}_i=O(1)$, 
if the corresponding operator is not suppressed by some 
protective symmetry (such as $B$ or $L$ conservation).
The observation that this expectation is {\em not} 
fulfilled by several dimension-six operators contributing 
to flavour-changing processes is often denoted 
as the {\em flavour problem}.

Let's consider for instance the following set of 
$\Delta F=2$ dimensions-six operators
\be
\cO^{\Delta F=2}_{ij} = (\bar Q_L^i \gamma^\mu Q_L^j )^2 ~,
\qquad Q^i_L =\left(\ba{c} u^i_L \\ d^i_L \ea \right)~,
\label{eq:dfops}
\ee
where $i,j=1\ldots 3$ are flavour indices and, 
for reasons that will become clear in the following, 
we work in mass-eigenstate basis of down-type quarks.
These operators contribute 
at the tree-level to the meson-antimeson mixing amplitudes,
which in the SM are generated only at the one-loop level. 
The case of $K^0$--$\bar K^0$, 
$B_d$--$\bar B_d$, $B_s$--$\bar B_s$ mixing
is particularly interesting: here the SM contribution is 
dominated by the short-distance contribution of top-quark 
loops 
\be
 \cA^{\Delta F=2}_{\rm SM} \approx  
   \frac{ m_t^2 }{16 \pi^2 v^4} \left(V_{3i}^* V_{3j} \right)^2\times 
    \langle \bar  M |  (\bar d_L^i \gamma^\mu d_L^j )^2  
| M \rangle \qquad [M = K^0, B_d, B_s]~,
\ee
($V_{ij}$ denote the elements of the CKM matrix)
and can be computed with high accuracy.
Moreover, both moduli and phases of these
three mixing amplitudes have been determined 
with good accuracy  from experiments
(with the exception of the
CP-violating case in $B_s$--$\bar B_s$ mixing).
In all cases the magnitude of the new-physics amplitude
(or the term generated by the dimension-six operators) 
cannot exceed, in size, the SM short-distance contribution.
Denoting $c_{ij}$ the couplings of the non-standard 
operators in (\ref{eq:dfops}), the condition 
$| \cA^{\Delta F=2}_{\rm NP}| <  | \cA^{\Delta F=2}_{\rm SM} |$
implies
\bea
\Lambda < \frac{ 3.4~{\rm TeV} }{| V_{3i}^* V_{3j}|/|c_{ij}|^{1/2}  }
<  \left\{ \ba{l}  
9\times 10^3~{\rm TeV} \times |c_{21}|^{1/2} \qquad {\rm from} \quad 
K^0-\bar K^0
 \\
4\times 10^2~{\rm TeV} \times |c_{31}|^{1/2} \qquad {\rm from}  \quad
B_d-\bar B_d
 \\
7\times 10^1~{\rm TeV} \times |c_{32}|^{1/2} \qquad {\rm from}  \quad
B_s-\bar B_s \ea
\right. 
\label{eq:boundsDF2}
\eea 

The message of these bounds is quite clear: if we want 
to keep $\Lambda$ in the TeV range,  the EFT must have 
a highly non-generic flavour structure.  
In the specific
case of the $\Delta F=2$ operators in  (\ref{eq:dfops}),
we must find a symmetry argument such that  
$|c_{ij}| \lsim  |V_{3i}^* V_{3j}|$.
A similar problem is found also in the case of 
$\Delta F=1$ operators contributing to flavour-changing
neutral-current (FCNC) processes. In this case the 
bounds are different, but the main problem is the same:
FCNC and $\Delta F=2$ amplitudes are suppressed in the SM not only 
by the typical $1/(4\pi)^2$ of loop amplitudes, 
but also by the GIM mechanisms and the hierarchy of the 
CKM matrix ($|V_{3i}|\ll 1$, for $i\not =3$). 

The most reasonable (but also most {\em pessimistic}) 
solution to the flavour problem is the so-called 
Minimal Flavour Violation hypothesis
that will be discussed below.

\section{Minimal Flavour Violation in the quark sector}

The main idea of MFV is that flavour-violating 
interactions are linked to the
known structure of Yukawa couplings also beyond the SM. 
In a more quantitative way, the MFV construction consists 
in identifying the flavour symmetry and symmetry-breaking structure 
of the SM and enforce it in the EFT. 

The largest group of flavour-changing field 
transformations commuting with the SM gauge group is 
${\mathcal G}_{f} = {\mathcal G}_{q} \otimes {\mathcal G}_{\ell} \otimes U(1)^5$, where 
\beq
{\mathcal G}_{q}
= {\rm SU}(3)_{Q_L}\otimes {\rm SU}(3)_{U_R} \otimes {\rm SU}(3)_{D_R},
\qquad 
{\mathcal G}_{\ell}
=  {\rm SU}(3)_{L_L} \otimes {\rm SU}(3)_{E_R}
\label{eq:Ggroups}
\eeq
and three of the five $U(1)$ charges can be identified with 
baryon number, lepton number and 
hypercharge~\cite{Georgi}.\footnote{Since hypercharge is gauged 
and involves also the Higgs field, it may be more convenient not 
to include it in the flavour group, which would then 
be defined as ${\mathcal G}_{\rm SM} = {\mathcal G}_{\ell} 
\otimes U(1)^4$~\cite{Feldmann:2009dc}.}
This large group and, particularly the ${\rm SU}(3)$ 
subgroups controlling flavour-changing transitions, is 
explicitly broken by the Yukawa interaction
\beq
\cL_Y  =   {\bar Q}_L \yuk_d D_R  H
+ {\bar Q}_L {\yuk_u} U_R  H_c
+ {\bar L}_L {\yuk_e} E_R  H {\rm ~+~h.c.}
\label{eq:LY}
\eeq
Since ${\mathcal G}_{f}$ is broken already within the SM, 
it would not be consistent to impose it as an exact symmetry 
of the additional degrees of freedom
present in SM extensions: even if absent a the tree-level,
the breaking of ${\mathcal G}_{f}$ would reappear at the quantum level 
because of the Yukawa interaction.  
The most restrictive hypothesis 
we can make to {\em protect} the breaking of ${\mathcal G}_{f}$ 
in a consistent way, is to assume that 
$\yuk_d$, $\yuk_u$ and $\yuk_e$ are the only source of 
${\mathcal G}_{f}$-breaking also beyond the SM.

To derive the phenomenological consequences of this hypothesis, 
it is convenient to treat ${\mathcal G}_{f}$ as an unbroken 
symmetry of the underlying theory, promoting  the $\yuk_i$ to be 
non-dynamical fields (spurions) with non-trivial transformation properties 
under ${\mathcal G}_{f}$
\beq
\yuk_u \sim (3, \bar 3,1)_{{\rm SU}(3)^3_q}~,\qquad
\yuk_d \sim (3, 1, \bar 3)_{{\rm SU}(3)^3_q}~,\qquad 
\yuk_e \sim (3, \bar 3)_{{\rm SU}(3)^2_\ell}~.
\eeq
Employing the EFT language, we then define that an effective theory satisfies 
the criterion of Minimal Flavour Violation if all higher-dimensional operators,
constructed from SM fields and $\yuk$ spurions, are 
invariant under the flavour group 
${\mathcal G}_{f}$~\cite{D'Ambrosio:2002ex}.

According to this criterion, one should in principle 
consider operators with arbitrary powers of the (adimensional) 
Yukawa spurions. However, a strong simplification arises by the 
observation that all the eigenvalues of the Yukawa matrices 
are small, but for the top one, and that the off-diagonal 
elements of the CKM matrix ($V_{ij}$) are very suppressed.
In the mass-eigenstate basis of down-type quarks, the 
two quark Yukawa coupling assume the following form:
\be
\yuk_d = 
{\rm diag}(y_d,y_s,y_b)~, \qquad
\yuk_u =  V^\dagger \times {\rm diag}(y_u,y_c,y_t)~, \qquad
y_i = \frac{\sqrt{2} m_i}{v}~.
\ee
It is then easy to realize that, similarly to the pure SM case, 
the leading coupling ruling all FCNC transitions 
with external down-type quarks is \cite{D'Ambrosio:2002ex}
\beq
(\Delta^q_{\rm LL})_{i\not=j} =  (\yuk_u \yuk_u^\dagger)_{ij}
\approx y_t^2  V^*_{3i} V_{3j}~,
\qquad 
  y_t =m_t/v \approx 1~.
\eeq 
Higher-order spurion combinations contributing to FCNCs
are either negligible or proportional to $\Delta^q_{\rm LL}$.

The suppression of the off-diagonal entries of $\Delta^q_{\rm LL}$
implies that, within the MFV framework, 
the bounds on the scale of dimension-six FCNC 
effective operators are in the few TeV range (detailed 
bounds for $\Delta F=2$ and $\Delta F=1$ operators 
can be found in \cite{Bona:2007vi} and \cite{Hurth:2008jc},
respectively). Moreover, the flavour structure 
of $\Delta^q_{\rm LL}$ implies a well-defined link among 
possible deviations from the SM in FCNC transitions 
of the type $s\to d$, $b\to d$, and  
$b\to s$ (the only quark-level transitions where 
observable deviations from the SM are expected).

\subsection{General considerations}
  
The idea that the CKM matrix rules the strength of 
FCNC transitions also beyond the SM is a concept
that has been implemented and discussed in several works, 
especially after the first results of the $B$ factories
(see e.g.~Ref.~\cite{MFV2,Buras:2003jf}). 
However, it is worth stressing that the CKM matrix 
represents only one part of the problem: a key role in
determining the structure of FCNCs is also played  by quark masses
(via the GIM mechanism), or by the Yukawa eigenvalues. 
In this respect, the above MFV criterion provides the maximal protection 
of FCNCs (or the minimal violation of flavour symmetry), since the full 
structure of Yukawa matrices is preserved. Moreover, 
contrary to other approaches, the above MFV criterion 
is based on a renormalization-group-invariant symmetry argument,
which can easily be extended to TeV-scale 
effective theories where new degrees of freedoms, 
such as extra Higgs doublets or SUSY partners of the SM fields
are included. 

In particular, it is worth stressing that the MFV hypothesis 
can be implemented also if the EFT does not 
include a light Higgs boson in the spectrum. In this case 
Eq.~(\ref{eq:LY}) is replaced by an effective interaction 
between fermion fields and the Goldstone bosons associated 
to the spontaneous breaking of the gauge symmetry.
From the point of view of the flavour symmetry, this interaction 
is identical to the one in (\ref{eq:LY}), and this allows us 
to proceed as in  the case with the explicit Higgs filed 
(see e.g.~Ref.~\cite{Barbieri:2008zt}).
The only difference between weakly- and strongly-interacting 
theories at the TeV scale is that in the latter case the 
expansion in powers of the Yukawa spurions is not 
necessarily a rapidly convergent series. If this is the case, 
then a resummation of the terms involving the top-quark Yukawa 
coupling needs to be performed~\cite{Feldmann:2008ja,Kagan:2009bn}.

\begin{figure}[t]
\begin{center}
\includegraphics[width=65mm]{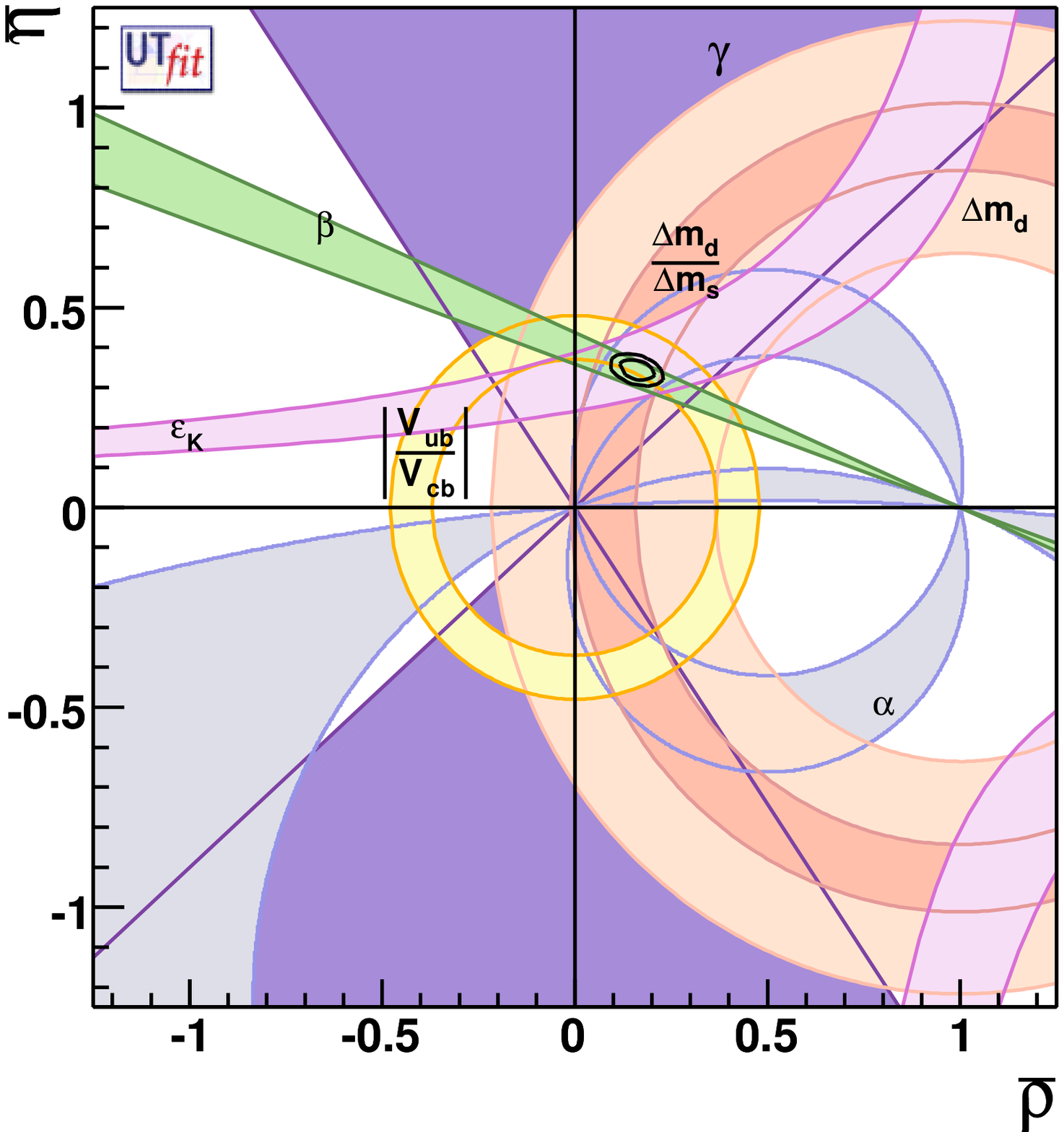}
\includegraphics[width=65mm]{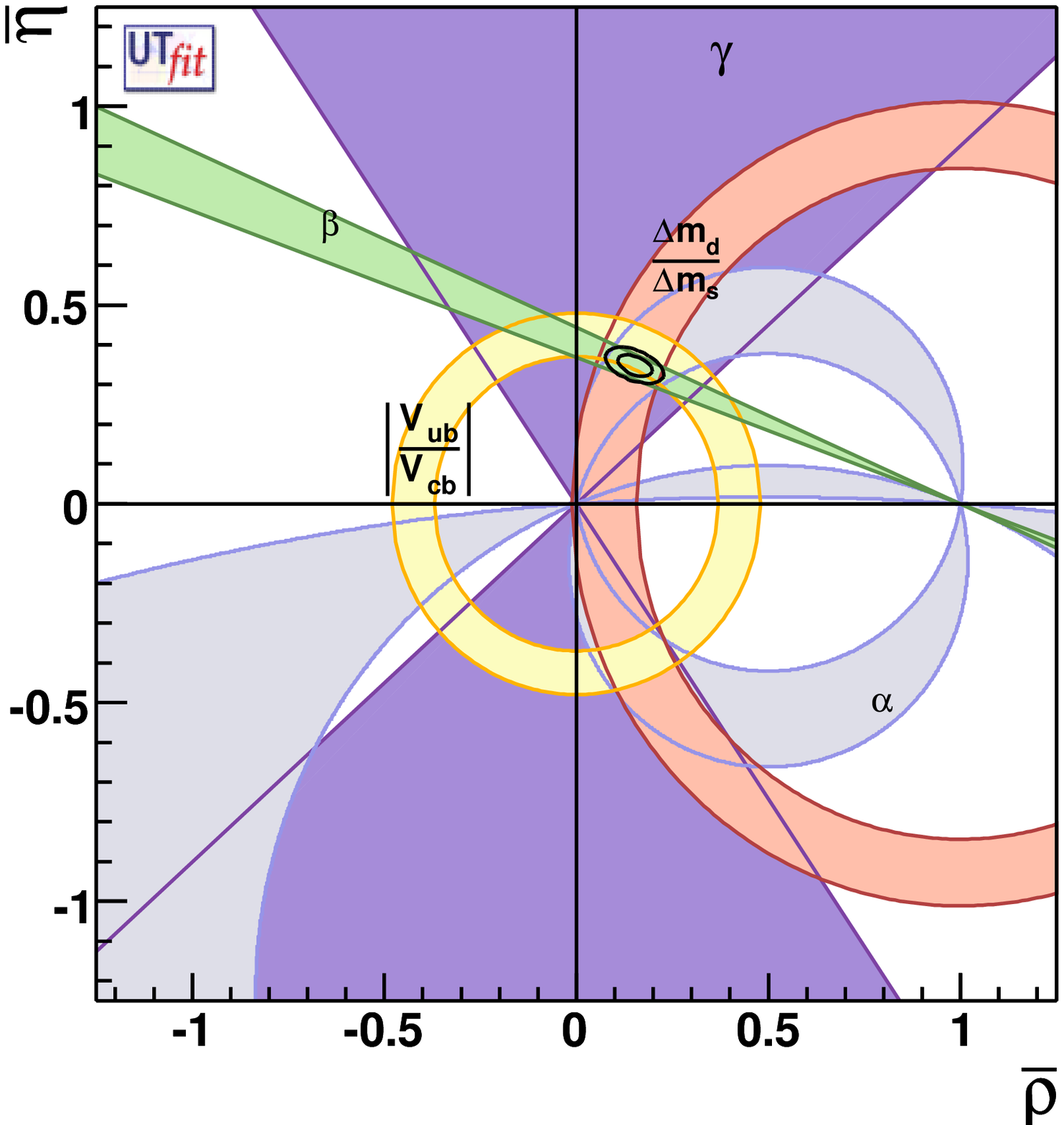}
\caption{\label{fig:UTfits} Fit of the CKM unitarity 
triangle (in 2008) within the SM (left) and 
in generic extensions of the SM 
satisfying the MFV hypothesis (right)~\cite{Bona:2007vi}. }
\end{center}
\end{figure}

As shown in Fig.~\ref{fig:UTfits}, the MFV hypothesis provides a
natural (a posteriori) justification of why no NP effects have 
been observed in the quark sector: by construction, most of the clean 
observables measured at $B$ factories are insensitive to NP effects 
in this framework. However, it should be stressed that we are still
very far from having proved the validity of this hypothesis 
from data. A proof of the MFV hypothesis can be achieved 
only with a positive evidence of physics beyond 
the SM exhibiting the flavour pattern 
predicted by the MFV assumption.
This proof could come from the $B_{s,d}\to \ell^+\ell^-$ or 
the $K\to \pi\nu\bar\nu$ systems: in both cases there
is still room for sizable non-standard contributions 
(even within the MFV framework), and we can identify two 
observables unambiguously correlated by the MFV 
hypothesis~ \cite{Hurth:2008jc,Blanke:2006ig}

This model-independent structure does not hold in 
most of the alternative definitions of MFV models 
that can be found in the literature. For instance, 
the definition of Ref.~\cite{Buras:2003jf} 
(denoted constrained MFV, or CMFV)
contains the additional requirement that the 
effective FCNC operators playing a significant 
role within the SM are the only relevant ones 
also beyond the SM. 
This condition is realized only in weakly coupled 
theories at the TeV scale with only one light Higgs doublet, 
such as the MSSM with small $\tan\beta$.
It does not hold in several other frameworks, such as 
Higgsless models, or the MSSM with large $\tan\beta$.

\subsection{MFV at large $\tan\beta$.}
\label{eq:largetanb}

If the Yukawa Lagrangian contains more than one Higgs field,
we can still assume that the Yukawa couplings are the only 
irreducible breaking sources of ${\mathcal G}_{q}$, but 
we can change their overall normalization.
A particularly interesting scenario
is the two-Higgs-doublet model where 
the two Higgses are coupled separately to up-
and down-type quarks:
\begin{equation}
\mathcal{L}^{2HDM}_{Y}  =   {\bar Q}_L \lambda_d D_R  H_D
+ {\bar Q}_L \lambda_u U_R  H_U
+ {\bar L}_L \lambda_e E_R  H_D {\rm ~+~h.c.}
\label{eq:LY2}
\end{equation}
This Lagrangian is invariant under an extra ${\rm U}(1)$
symmetry with respect to the one-Higgs Lagrangian 
in Eq.~(\ref{eq:LY}): a symmetry under which 
the only charged fields are $D_R$ and $E_R$ 
(charge $+1$) and  $H_D$ (charge $-1$).
This symmetry, denoted ${\rm U}_{\rm PQ}$, 
prevents tree-level FCNCs and implies that $\lambda_{u,d}$ are the only 
sources of ${\mathcal G}_{q}$ breaking appearing in the Yukawa
interaction (similar to the one-Higgs-doublet
scenario). Coherently with the MFV hypothesis, we can then 
assume that $\lambda_{u,d}$ are the only relevant 
sources of ${\mathcal G}_{q}$ breaking appearing 
in all the low-energy effective operators. 
This is sufficient to ensure that flavour-mixing 
is still governed by the CKM matrix, and naturally guarantees
a good agreement with present data in the $\Delta F =2$
sector. However, the extra symmetry of the Yukawa interaction allows
us to change the overall normalization of 
$\lambda_{u,d}$ with interesting phenomenological consequences
in specific rare modes. 

The normalization of the  Yukawa couplings is controlled
by the ratio of the vacuum expectation values  of the two Higgs fields, 
or by the parameter $\tan\beta = \langle H_U\rangle/\langle H_D\rangle$.
For $\tan\beta >\!\! > 1 $ the smallness of the $b$ quark 
and $\tau$ lepton masses can be attributed to the smallness 
of $1/\tan\beta$ rather than to the corresponding Yukawa couplings.
As a result, for $\tan\beta >\!\! > 1$ we cannot anymore neglect 
the down-type  Yukawa coupling. 
Moreover, the ${\rm U}(1)_{\rm PQ}$ symmetry cannot be exact:
it has to be broken at least in the scalar potential
in order to avoid the presence of a massless pseudoscalar Higgs.
Even if the breaking of  ${\rm U}(1)_{\rm PQ}$  and 
 ${\mathcal G}_{q}$ are decoupled, the presence of 
${\rm U}(1)_{\rm PQ}$ breaking sources can have important 
implications on the  structure of the Yukawa interaction,
especially if $\tan\beta$ is large~\cite{Hall:1993gn,Blazek:1995nv,
Babu:1999hn,Isidori:2001fv}.
We can indeed consider new dimension-four operators such as
\begin{equation}
 \epsilon~ {\bar Q}_L  \lambda_d D_R  (H_U)^c
\qquad {\rm or} \qquad 
 \epsilon~ {\bar Q}_L  \lambda_u\lambda_u^\dagger \lambda_d D_R  (H_U)^c~,
\label{eq:O_PCU}
\end{equation}
where $\epsilon$ denotes a generic MFV-invariant
${\rm U}(1)_{\rm PQ}$-breaking source. Even if $\epsilon \ll 1 $, 
the product $\epsilon \times  \tan\beta$ can be $\mathcal{O}(1)$, 
inducing large corrections to the down-type Yukawa 
sector:
\begin{equation}
 \epsilon~ {\bar Q}_L  \lambda_d D_R  (H_U)^c 
\ \stackrel{vev}{\longrightarrow}  \
 \epsilon~  {\bar Q}_L  \lambda_d D_R   \langle H_U \rangle = 
  (\epsilon\times\tan\beta)~  {\bar Q}_L  \lambda_d D_R   \langle H_D \rangle~.
\end{equation}

Since the $b$-quark Yukawa coupling becomes $\mathcal{O}(1)$, 
the large-$\tan\beta$ regime is particularly interesting 
for helicity-suppressed observables in $B$ physics. 

One of the clearest phenomenological 
consequences is a suppression (typically in the $10-50\%$ range)
of the $B \to \ell \nu$ decay
rate with respect to its SM expectation~\cite{Hou:1992sy,akeroyd,Isidori:2006pk}.
But the most striking signature could arise from the 
rare decays $B_{s,d}\to \ell^+\ell^-$
whose rates could be enhanced over the SM expectations 
by more than one order of magnitude~\cite{Hamzaoui:1998nu,Choudhury:1998ze,Babu:1999hn}.
An enhancement of both $B_{s}\to \ell^+\ell^-$ and 
$B_{d}\to \ell^+\ell^-$ respecting the MFV relation
$\Gamma(B_{s}\to \ell^+\ell^-)/\Gamma(B_{d}\to \ell^+\ell^-)
\approx |V_{ts}/V_{td}|^2$ would be an unambiguous signature 
of MFV at large $\tan\beta$~\cite{Hurth:2008jc}.

\subsection{MFV in supersymmetry}

One of the explicit new-physics framework where the MFV hypothesis 
could be more easily implemented is the minimal supersymmetric
extension of the SM (MSSM). Since the squark fields have 
well-defined transformation
properties under the SM quark-flavour group ${\mathcal G}_q$,
the MFV hypothesis can easily be implemented in the MSSM 
framework following the general rules outlined above.

We need to consider all possible interactions compatible 
with i) softly-broken supersymmetry; ii) the breaking of 
${\mathcal G}_q$ via the spurion fields $\lambda_{u,d}$. 
This allows to express the squark mass terms and 
the trilinear quark-squark-Higgs couplings 
as follows~\cite{Hall:1990ac,D'Ambrosio:2002ex}:
\begin{eqnarray}
{\tilde m}_{Q_L}^2 &=& {\tilde m}^2 \left( a_1 {1 \hspace{-.085cm}{\rm l}} 
+b_1 \lambda_u \lambda_u^\dagger +b_2 \lambda_d \lambda_d^\dagger 
+b_3 \lambda_d \lambda_d^\dagger \lambda_u \lambda_u^\dagger
+b_4 \lambda_u \lambda_u^\dagger \lambda_d \lambda_d^\dagger +\ldots
 \right)~, \nonumber  \\
{\tilde m}_{U_R}^2 &=& {\tilde m}^2 \left( a_2 {1 \hspace{-.085cm}{\rm l}} 
+b_5 \lambda_u^\dagger \lambda_u +\ldots \right)~,  \qquad 
A_U ~=~ A\left( a_3 {1 \hspace{-.085cm}{\rm l}} 
+b_6 \lambda_d \lambda_d^\dagger +\ldots \right) \lambda_u~,\qquad 
\label{eq:MSSMMFV}
\end{eqnarray}
and similarly for the down-type terms. 
The dimensionful parameters $\tilde m$ and $A$,
expected to be in the range few 100 GeV -- 1 TeV, 
set the overall scale of the soft-breaking terms.
In Eq.~(\ref{eq:MSSMMFV}) we have explicitly shown  
all independent flavour structures which cannot be absorbed into 
a redefinition of the leading terms (up to tiny contributions 
quadratic in the Yukawas of the first two families),  
when $\tan\beta$ is not too large and the bottom Yukawa coupling 
is small, the terms quadratic in $\lambda_d$ can be dropped.

In a bottom-up approach, the dimensionless coefficients 
$a_i$ and $b_i$ should be considered as free parameters 
of the model. Note that  this structure is 
renormalization-group invariant: the values of 
$a_i$ and $b_i$ change according to the 
Renormalization Group (RG) flow, but the general structure 
of Eq.~(\ref{eq:MSSMMFV})
is unchanged. This is not the case if the $b_i$ are set to zero,
corresponding to the so-called hypothesis of {\em flavour universality}.
In several explicit mechanisms of supersymmetry breaking, 
the condition of flavour universality
holds at some high scale $M$, such as the scale of  
Grand Unification  or the mass-scale of the
messenger particles in Gauge Mediation (see Ref.~\cite{Giudice:1998bp}). 
In this case  non-vanishing 
$b_i \sim (1/4\pi)^2 \ln M^2/ {\tilde m}^2$ are 
generated by the RG evolution. 
As recently 
pointed out in Ref.~\cite{Paradisi:2008qh,Colangelo:2008qp}, 
the RG flow in the MSSM-MFV 
framework exhibits quasi infra-red fixed points: even
if we start with all the $b_i =\mathcal{O}(1)$ at some high scale, 
the only non-negligible terms at the TeV scale are those 
associated to the $\lambda_u \lambda_u^\dagger$ structures.

While MFV can easily be implemented in the MSSM, 
it is worth to stress that the flavour problem could 
have a different solution in supersymmetric extensions of the SM.
For instance,  
one interesting possibility is that there is no special 
flavor symmetry symmetry, but the first two generations 
of squarks are well above the TeV scale~(see Ref.~\cite{Giudice:2008uk}
for a recent analysis). Keeping only the third generation light
is sufficient to stabilise the Higgs sector, and  
with heavy squarks for the first two 
generations the severe bound from the kaon system 
in (\ref{eq:boundsDF2}) are less problematic.

\section{MFV in the lepton sector}
Apart from arguments based on the analogy between quarks
and leptons, the introduction of a MFV hypothesis 
for the lepton sector (MLFV) is demanded 
by a severe fine-tuning problem also in the lepton sector:
within a generic EFT approach, the non-observation of 
$\mu\rightarrow e\gamma$ implies an effective NP scale 
above $10^5$ TeV unless the coupling of the corresponding 
effective operator is suppressed by some symmetry principle. 

Since the observed neutrino mass parameters are not 
described by the SM Yukawa interaction in Eq.~(\ref{eq:LY}),
the formulation of a MLFV hypothesis is not straightforward,
and some additional hypothesis is needed. 
A first natural assumption is that  
the breaking of the total lepton number ($L$) and the 
lepton flavour --the ${\mathcal G}_{\ell}$
group in (\ref{eq:Ggroups})--
are decoupled in the underlying theory. 
Following Ref.~\cite{MLFV} we can then  
consider two main scenarios. 
They are characterized by the different status assigned to
the effective Majorana mass matrix $g_\nu$  appearing as 
coefficient of the $L$-violating  dimension-five operator in the low 
energy effective theory~\cite{Weinberg:1979sa}:
\beq
\cL_{\rm eff}^{\nu}  = -\frac{1}{ \Lambda_{\rm LN}}\,g_\nu^{ij}(\bar
L^{ci}_L\tau_2 H)(H^T\tau_2L^j_L)  {\rm ~+~h.c.} \quad \longrightarrow 
\quad m_\nu =  \frac{g_\nu v^2}{\Lambda_{\rm LN}}
\eeq
In the truly minimal scenario (dubbed {\em minimal field content}),  
$g_\nu$ and the charged-lepton
Yukawa coupling ($\lambda_e$) are assumed to be the only irreducible sources 
of breaking of ${\mathcal G}_{\ell}$, the lepton-flavour 
symmetry of the low-energy theory. 

The irreducible character of $g_\nu$
does not hold in many realistic underlying 
theories with heavy right-handed neutrinos. For this reason, 
a second scenario (dubbed {\em extended field content}), 
with heavy right-handed neutrinos and a larger 
lepton-flavour symmetry group, ${\mathcal G}_{\ell}\times {\rm O}(3)_{\nu_R}$,
has also been considered. 
In this extended scenario, a natural and 
economical choice about the symmetry-breaking terms is the identification 
of the two Yukawa couplings, $\lambda_\nu$ and $\lambda_e$, 
as the only irreducible symmetry-breaking structures.
In this context,
$g_\nu \sim  \lambda_\nu^T \lambda_\nu$ and the 
$L$-violating  mass term of the heavy 
right-handed neutrinos is flavour-blind 
(up to Yukawa-induced corrections):
\beqa
\cL_{\rm heavy}  &=&  -\frac{1}{2} M_\nu^{ij}\bar \nu^{ci}_R\nu_R^j {\rm ~+~h.c.} 
\qquad M_\nu^{ij}=M_\nu  \delta^{ij}  \no\\
\cL^{\rm ext}_{Y}  &=&  \cL_{Y}  +i \lambda_\nu^{ij}\bar\nu_R^i(H^T \tau_2L^j_L) {\rm ~+~h.c.} 
\eeqa
In this scenario the flavour changing coupling 
relevant to $l_i\rightarrow l_j\gamma$ decays reads
\begin{eqnarray}
\label{eq:MVFDRLext}
(\Delta^{\ell}_{\rm LR} )_{\rm MLFV} ~ \propto ~ \lambda_e \lambda_\nu^\dagger
\lambda_\nu 
 \to  \frac{m_e}{v}\, \frac{M_\nu}{v^2}U_{\rm PMNS}\, (m^{1/2}_{\nu})_{\rm diag} H^2
(m^{1/2}_{\nu})_{\rm diag}\,
U^\dagger_{\rm PMNS}
\end{eqnarray} 
where  $H$ is an Hermitian-orthogonal matrix  which can
be parametrized in terms of three real parameters ($\phi_i$) which
control the amount of CP-violation in the right-handed 
sector~\cite{Pascoli:2003rq}.
In the CP-conserving limit, $H \rightarrow I$ and the 
phenomenological predictions for lepton FCNC decays 
turns out to be quite similar to the minimal
field content scenario~\cite{MLFV}. 

Once the field content of model is extended, there are in principle many 
alternative options to define the irreducible sources of lepton flavour 
symmetry breaking (see Ref.~\cite{Davidson:2006bd,Hambye} 
for an extensive discussion and alternative scenarios). 
The specific choice discussed above has two important advantages:
it is predictive and closely resemble the MFV hypothesis in the quark sector. 
The $\nu_R$'s are the counterpart of right-handed up quarks and, similarly to 
the quark sector, the symmetry-breaking sources are two Yukawa couplings.

The basic assumptions of the MLFV hypotheses 
are definitely less data-driven 
with respect to the quark sector. Nonetheless, the formulation of 
an EFT based on these assumptions is still very useful.
As I will briefly illustrate in the following, it allows us to 
address in a very general way the following fundamental 
question: how can we detect the presence of new irreducible 
(fundamental) sources of LF symmetry breaking?

\subsection{Phenomenological consequences}

\begin{figure}[t]
\begin{center}
\includegraphics[width=150mm]{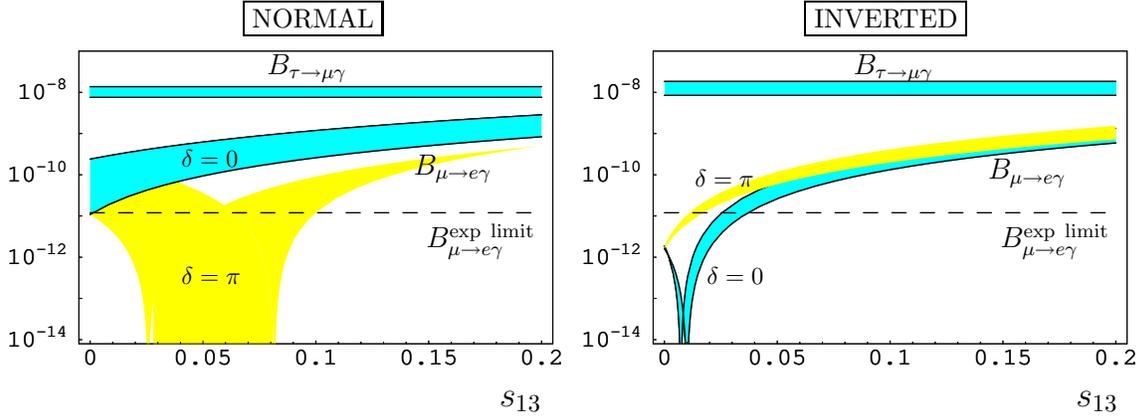}
\caption{\label{fig:MLFV}  
$B_{\tau \to  \mu \gamma} = \Gamma(\tau \to \mu \gamma)/\Gamma(\tau \to \mu \nu \bar{\nu}) $
compared to the $\mu  \to  e \gamma$ constraint within MLFV (minimal field
content),  as a function of the neutrino mixing angle $s_{13}$~\cite{MLFV}.
The shading corresponds to different values of the phase $\delta$ 
and the normal/inverted spectrum. The NP scales have been set to
 $\Lambda_{\rm LN}/\Lambda = 10^{10}$; their variation affects 
only the overall vertical scale.}
\end{center}
\end{figure}

Using the MLFV-EFT approach, one can easily 
demonstrate that --in absence of new sources of LF violation--
visible FCNC decays of $\mu$ and $\tau$ can occur only 
if there is a large hierarchy between $\LLFV$
(the scale of  new degrees of freedoms carrying LF) 
and $\LLN \sim M_\nu$ (the scale of total LN violation)~\cite{MLFV}. 
More interestingly, the EFT  allows us to draw unambiguous 
predictions about the relative size of LF violating decays 
of charged leptons (in terms of neutrino masses and mixing angles). 
At present, the uncertainty in the predictions for such ratios is limited 
from the poorly constrained value of the $1$--$3$ mixing angle in the neutrino
mass matrix ($s_{13}$) and, to a lesser extent, 
from the neutrino spectrum ordering and the CP violating phase $\delta$. 
One of the more clear consequences from the phenomenological
point of view is the observation that  if $s_{13} \gsim 0.1$ there 
is no hope to observe $\tau \to \mu \gamma$ at future accelerators 
(see Fig.~\ref{fig:MLFV}). This happens because the stringent
constraints from $\mu \to e \gamma$ already forbid too low values 
for the effective scale of LF violation. In other words,
in absence of new sources of LF violation the most sensitive 
FCNC probe in the lepton sector is $\mu \to e \gamma$.
This process should indeed be observed at  MEG~\cite{MEG}
for very realistic values of the new-physics scales $\LLFV$ 
and $\LLN$.  Interestingly enough, this conclusion holds 
both in the minimal- and in the extended-field-content
formulation of the MLFV framework. 

The expectation of a higher NP sensitivity of 
$\mu \to \mu \gamma$ with respect to $\tau \to \mu \gamma$
(taking into account the corresponding experimental resolutions) 
is confirmed in several realistic NP frameworks.
This happens for instance in the MSSM scenarios 
analysed in Ref.~\cite{hisano,profumo,Herrero}
with the exception of specific corners 
of the parameter space~\cite{hisano}.

In the MLFV scenario with extended field content we can 
generate the observed matter-antimatter asymmetry of the 
Universe by means of leptogenesis~\cite{Yanagida}. The viability 
of leptogenesis within the MLFV framework, which has been
demonstrated in Ref.~\cite{Cirigliano:2006nu,Branco:2006hz,Cirigliano:2007hb},
is an interesting conceptual point: it implies that there are no 
phenomenological motivations to introduce new sources of flavour symmetry 
breaking in addition to the four $\lambda_i$ (the three SM Yukawa couplings 
and  $\lambda_\nu$). A necessary condition for leptogenesis to occur is a 
non-degenerate heavy-neutrino spectrum. Within the MLFV
framework, the tree-level degeneracy of heavy neutrinos 
is lifted only by radiative corrections, which implies a rather
predictive/con\-stra\-ined scenario.
From the phenomenological point of view, an important difference
with respect to the CP-con\-ser\-ving case is the fact that 
non-vanishing $\phi_i$ change the predictions of the LFV decays, 
typically producing a further 
enhancement of the  $\mathcal{B}(\mu\to e\gamma)/\mathcal{B}(\tau\to \mu\gamma)$
ratio~\cite{Branco:2006hz}.

\section{Beyond Minimal Flavour Violation}

Despite its phenomenological success, 
there are various reasons to expect some deviations
from the MFV hypothesis. In the following we illustrate 
two well-motivated examples where such deviations are
indeed expected. 

Interestingly, in both cases the 
phenomenological signatures of such deviations are 
expected in the kaon system rather than in the $B$ system. 
This is not surprising given the strong suppression of 
short-distance $s\to d$ FCNC transitions in the SM.

\subsection{Grand Unified Theories}
\label{sec:MFVgut}
Once we accept the idea that flavour dynamics obeys a MFV
principle, both in the quark and in the lepton sector, it is
interesting to ask if and how this is compatible with a
grand-unified theory (GUT), where quarks and leptons sit in the same
representations of a unified gauge group. This question has 
been addressed in Ref.~\cite{Grinstein:2006cg}, 
considering the exemplifying case of ${\rm SU}(5)_{\rm gauge}$.

Within ${\rm SU}(5)_{\rm gauge}$, the down-type singlet quarks ($d^c_{iR}$) and the lepton doublets 
($L_{iL}$) belong to the $\bar {\bf 5}$ representation; the quark doublet
($Q_{iL}$), the up-type ($u^c_{iR}$) and lepton singlets ($e_{iR}^c$) 
belong to the ${\bf 10}$ representation, and finally 
the right-handed neutrinos ($\nu_{iR}$) are singlet.
In this framework the largest 
group of flavour transformation commuting with the gauge group is
${\mathcal G}_{\rm GUT} = {\rm SU}(3)_{\bar 5} \times {\rm SU}(3)_{10}\times {\rm SU}(3)_1$, 
which is smaller than the direct product 
of the quark and lepton groups discussed before 
(${\mathcal G}_q \times {\mathcal G}_l$).  
We should therefore expect some violations of the MFV+MLFV predictions,
either in the quark sector, or in the lepton sector, or in both. 

A phenomenologically acceptable description of the low-energy fermion 
mass matrices requires the introduction of at least four irreducible 
sources of ${\mathcal G}_{\rm GUT}$ breaking. From this point of view
the situation is apparently similar to the non-unified case: the four 
 ${\mathcal G}_{\rm GUT}$ spurions can be put in one-to-one 
correspondence with the low-energy spurions $\lambda_u$,$\lambda_d$, 
$\lambda_e$, and $\lambda_\nu$. However, the smaller flavour group
does not allow the diagonalization of $\lambda_d$ and
$\lambda_e$ (which transform in the same way under ${\mathcal G}_{\rm GUT}$)
in the same basis. As a result, two additional mixing matrices 
can appear in the expressions of flavour-changing rates~\cite{Grinstein:2006cg}.
The hierarchical texture of the new mixing matrices is known
since they reduce to the identity matrix in the limit 
$\lambda_e^T = \lambda_d$. Taking into account this fact, and
analysing the structure of the allowed higher-dimensional operators, 
a number of reasonably  firm phenomenological consequences  
can be deduced~\cite{Grinstein:2006cg}: 
\begin{itemize}
\item  
There is a well defined limit in which the standard MFV scenario 
for the quark  sector is  fully recovered: $M_\nu \ll 10^{12}$ GeV
(hence sufficiently small neutrino Yukawa couplings) 
and small $\tan \beta$ (in a two-Higgs doublet case). 
For $M_\nu \sim  10^{12}$ GeV and small $\tan \beta$, deviations from the standard MFV pattern 
can be expected in rare $K$ decays but  not in $B$ physics.
Ignoring fine-tuned 
scenarios, $M_\nu \gg  10^{12}$~GeV is excluded by the present constraints 
on quark FCNC transitions. Independently from the value of $M_\nu$, 
deviations from the standard MFV pattern can appear both in $K$ and in $B$ physics
for $\tan\beta~\gsim~m_t/m_b$. 
\item 
Contrary to the non-GUT MFV framework,  
the rate for $\mu \to e \gamma$ (and other LFV decays) cannot be 
arbitrarily suppressed by lowering the  average mass $M_\nu$ of the heavy  $\nu_R$.   
This fact can easily be understood by looking at the flavour structure 
of the relevant effective couplings, which now assume the following form:
\begin{equation}
\label{eq:MFVgut}
(\Delta^{\ell}_{\rm LR} )_{\rm MFV-GUT} = ~ c_{1}~\lambda_e
\lambda_\nu^\dagger\lambda_\nu ~+~c_{2}~\lambda_u
\lambda_u^\dagger\lambda_e ~+~c_{3}~\lambda_u
\lambda_u^\dagger\lambda_d^T\, +~\ldots
\end{equation}
In addition to the terms involving $\lambda_\nu\propto \sqrt{M_\nu}$ already 
present in the non-unified case, the GUT group allows also $M_\nu$-independent 
terms involving the quark Yukawa couplings. The latter become competitive 
for $M_\nu \lsim 10^{12}$ GeV and their contribution is such that for 
$\Lambda \lsim 10$ TeV  the  $\mu \to e \gamma$ rate is above 
$10^{-13}$ (i.e.~within the reach of  MEG~\cite{MEG}).
\item 
Improved experimental information on $\tau \to \mu \gamma$ and 
$\tau \to e \gamma$ are a now a key tool: the best observables to 
discriminate the relative size of the MLFV contributions with respect 
to the GUT-MFV ones. In particular, if the quark-induced terms turn out 
to be dominant, the  $\mathcal{B}(\tau\to\mu\gamma)/\mathcal{B}(\mu\to e\gamma)$
ratio could reach values of $\cO(10^{-4})$, allowing $\tau\to\mu\gamma$ 
to be just below the present exclusion bounds. 
\end{itemize}

\subsection{Flavour protection from hierarchical fermion profiles}
 
So far we have assumed that the suppression of flavour-changing
transitions beyond the SM can be attributed to a flavour symmetry, 
and a specific form of the symmetry-breaking terms. An interesting 
alternative is the possibility of a generic {\em  dynamical suppression}
of flavour-changing interactions, related to the weak mixing of 
the light SM fermions with the new dynamics at the TeV scale. 
A mechanism of this type is the so-called RS-GIM mechanism
occurring in models with a warped extra dimension.
In this framework the hierarchy of fermion masses, which 
is attributed  to the different localization of the fermions in 
the bulk~\cite{AS,RSoriginal}, implies that the lightest fermions 
are those localised far from the infra-red (SM) brane. 
As a result, the suppression of FCNCs involving 
light quarks is a consequence of the small overlap of the light 
fermions with the lightest Kaluza-Klein excitations~\cite{aps,Contino}. 

As pointed out in~\cite{Davidson:2007si}, 
also the general features of this class 
of models can be described by means of a general EFT 
approach. The two ingredients of this EFT are
the following: i) there exists a (non-canonical) basis 
for the SM fermions where their kinetic terms exhibit 
a rather hierarchical form: 
$$
 {\cal L}^{\rm quarks}_{\rm kin} = \sum_{\Psi=Q_L, U_R, D_R }
 \overline{\Psi} Z_\psi^{-2}   D\sla \Psi~,
\qquad Z_\psi =  {\rm diag}(z_\psi^{(1)}, z_\psi^{(2)}, z_\psi^{(3)}  )~, 
\qquad  z_\psi^{(1)}\ll z_\psi^{(2)}\ll z_\psi^{(3)} \lsim 1~,
$$
ii) in such basis there  is no flavour symmetry and all the 
flavour-violating interactions, including the Yukawa 
couplings, are $\cO(1)$. Once the fields are transformed 
into the canonical basis, the hierarchical kinetic terms 
act as a distorting lens, through which all interactions 
are seen as approximately aligned on the magnification axes 
of the lens. As anticipated, this construction provide
an effective four-dimensional description of a wide class of 
models with a with a warped extra dimension.
However, it should be stresses that this mechanism is not 
a general feature of models with extra dimensions: as discussed 
in~\cite{Cacciapaglia:2007fw,Csaki:2008eh}, 
also in extra-dimensional models is possible to postulate the 
existence of additional symmetries and, for instance, recover 
a MFV structure. On he other hand, hierarchical fermion profiles 
can be generated also in different theoretical frameworks: they could 
arise from Renormalisation Group running,
with large, positive, and distinct anomalous dimensions
for the different generations 
of fermions~\cite{Nelson:2000sn}.

The dynamical mechanism of hierarchical fermion profiles 
is quite effective in suppressing 
FCNCs beyond the SM. In particular, it can be shown 
that all the dimensions-six FCNC left-left operators
(such as the $\Delta F=2$ terms in (\ref{eq:dfops})), 
have the same suppression as in MFV~\cite{Davidson:2007si}.
However, a residual problem is present in the 
left-right operators contributing to CP-violating 
observables in the kaon 
system: $\epsilon_K$~\cite{Bona:2007vi,Csaki1,Blanke1}
(see also~\cite{Bauer:2008xb})
and  $\epsilon^\prime/\epsilon_K$~\cite{Davidson:2007si,noi}
(with potentially visible effects also in 
rare $B$ and $K$ decays~\cite{Blanke:2008yr}).
To suppress the latter, either the effective 
scale of new-physics is push up to $\sim 10$~TeV, 
or some form of alignment (of MFV type) must
be invoked (see e.g.~Ref.~\cite{Fitzpatrick:2007sa,Csaki:2009wc}
for a possible implementation of the alignment in the context
of models with warped extra dimensions).

\section{Conclusions} 
 
The absence of significant deviations from the SM 
in quark flavour physics is a key information about any 
extension of the SM.  Only models with a highly non 
generic flavour structure can both stabilise the 
electroweak sector and, at the same time, 
be compatible with flavour observables. 

A useful tool to identify the flavour structure 
of physics beyond the SM is provided by the 
construction of effective theories at the TeV scale, based on 
specific flavour symmetries and symmetry-breaking 
hypotheses, and compare them with data. 
As we have seen, the MFV hypothesis emerges 
as a natural candidate to {\em protect} flavour physics 
in extensions of the SM at the TeV scale. However, 
as discussed in the last chapter, MFV is    
not the only allowed possibility and is unlikely 
to be an exact symmetry principle. 

The identification of the flavour structure of 
physics beyond the SM remains an open issue.
Shedding more light on this problem require a twofold 
effort. On the one hand, we need to 
identify the TeV scale dynamics responsible for the 
breaking of the electroweak symmetry. On the other hand, 
we need more accurate measurements of clean 
FCNCs at low energies (such $B_{s,d} \to \ell^+\ell^-$, 
$K \to \pi \nu\bar\nu$, and the CP-violating phase of $B_s$ mixing)
to probe the favour symmetry-breaking pattern
of the new degrees of freedom.


\begin{thebibliography}{9}


\bibitem{Georgi}
 R.~S.~Chivukula and H.~Georgi, 
 \plb{188}{1987}{99}; 
 L.~J.~Hall and L.~Randall,
 Phys.\ Rev.\ Lett.\  {\bf 65}, 2939 (1990).

\bibitem{D'Ambrosio:2002ex}
  G.~D'Ambrosio, G.~F.~Giudice, G.~Isidori and A.~Strumia,
  Nucl.\ Phys.\ B {\bf 645} (2002) 155 
 [hep-ph/0207036].

\bibitem{Feldmann:2009dc}
  T.~Feldmann, M.~Jung and T.~Mannel,
  arXiv:0906.1523 [hep-ph].

\bibitem{Bona:2007vi}
  M.~Bona {\it et al.}  [UTfit Collaboration],
  JHEP {\bf 0803} (2008) 049
  [arXiv:0707.0636 [hep-ph]]; [http://www.utfit.org/].

\bibitem{Hurth:2008jc}
  T.~Hurth, G.~Isidori, J.~F.~Kamenik and F.~Mescia,
  Nucl.\ Phys.\  B {\bf 808} (2009) 326
  [arXiv:0807.5039 [hep-ph]].


\bibitem{MFV2}
 A.~Ali and D.~London, \epjc{9}{1999}{687}
 [hep-ph/9903535];
 A.~J.~Buras {\em et al.}, \plb{500}{2001}{161}
 [hep-ph/0007085];
 S.~Laplace, Z.~Ligeti, Y.~Nir and G.~Perez,
 \prd{65}{2002}{094040} [hep-ph/0202010].


\bibitem{Buras:2003jf}
  A.~J.~Buras,
  Acta Phys.\ Polon.\  B {\bf 34} (2003) 5615
  [arXiv:hep-ph/0310208].


\bibitem{Barbieri:2008zt}
  R.~Barbieri, G.~Isidori and D.~Pappadopulo,
  JHEP {\bf 0902} (2009) 029
  [arXiv:0811.2888 [hep-ph]].


\bibitem{Feldmann:2008ja}
  T.~Feldmann and T.~Mannel,
  Phys.\ Rev.\ Lett.\  {\bf 100} (2008) 171601
  [arXiv:0801.1802 [hep-ph]].


\bibitem{Kagan:2009bn}
  A.~L.~Kagan, G.~Perez, T.~Volansky and J.~Zupan,
  arXiv:0903.1794 [hep-ph].

\bibitem{Blanke:2006ig}
  M.~Blanke, A.~J.~Buras, D.~Guadagnoli and C.~Tarantino,
  JHEP {\bf 0610} (2006) 003
  [arXiv:hep-ph/0604057].




\bibitem{Hall:1993gn}
  L.~J.~Hall, R.~Rattazzi and U.~Sarid,
  Phys.\ Rev.\  D {\bf 50}, 7048 (1994)
  [hep-ph/9306309].

\bibitem{Blazek:1995nv}
  T.~Blazek, S.~Raby and S.~Pokorski,
  Phys.\ Rev.\  D {\bf 52}, 4151 (1995)
  [hep-ph/9504364].


\bibitem{Babu:1999hn}
  K.~S.~Babu and C.~F.~Kolda,
  Phys.\ Rev.\ Lett.\  {\bf 84}, 228 (2000)
  [hep-ph/9909476].

\bibitem{Isidori:2001fv}
  G.~Isidori and A.~Retico,
  JHEP {\bf 0111}, 001 (2001)
  [hep-ph/0110121].

\bibitem{Hou:1992sy}
  W.~S.~Hou,
  Phys.\ Rev.\  D {\bf 48}, 2342 (1993).

\bibitem{akeroyd}
A.~G.~Akeroyd and S.~Recksiegel,
J.\ Phys.\ G {\bf 29}, 2311 (2003)
[hep-ph/0306037].

\bibitem{Isidori:2006pk}
  G.~Isidori and P.~Paradisi,
  Phys.\ Lett.\  B {\bf 639}, 499 (2006)
  [hep-ph/0605012].



\bibitem{Hamzaoui:1998nu}
  C.~Hamzaoui, M.~Pospelov and M.~Toharia,
  Phys.\ Rev.\  D {\bf 59}, 095005 (1999)
  [hep-ph/9807350].

\bibitem{Choudhury:1998ze}
  S.~R.~Choudhury and N.~Gaur,
  Phys.\ Lett.\  B {\bf 451} (1999) 86
  [arXiv:hep-ph/9810307].





\bibitem{Hall:1990ac}
  L.~J.~Hall and L.~Randall,
  Phys.\ Rev.\ Lett.\  {\bf 65}, 2939 (1990).



\bibitem{Giudice:1998bp}
  G.~F.~Giudice and R.~Rattazzi,
  Phys.\ Rept.\  {\bf 322} (1999) 419
  [arXiv:hep-ph/9801271].

\bibitem{Paradisi:2008qh}
  P.~Paradisi, M.~Ratz, R.~Schieren and C.~Simonetto,
  Phys.\ Lett.\  B {\bf 668} (2008) 202
  [arXiv:0805.3989 [hep-ph]].


\bibitem{Colangelo:2008qp}
  G.~Colangelo, E.~Nikolidakis and C.~Smith,
  Eur.\ Phys.\ J.\  C {\bf 59} (2009) 75
  [arXiv:0807.0801 [hep-ph]].

\bibitem{Giudice:2008uk}
  G.~F.~Giudice, M.~Nardecchia and A.~Romanino,
  Nucl.\ Phys.\  B {\bf 813} (2009) 156
  [arXiv:0812.3610 [hep-ph]].





\bibitem{MLFV}
 V.~Cirigliano, B.~Grinstein, G.~Isidori and  M.~B.~Wise,
 \npb{728}{2005}{121} [hep-ph/0507001].



\bibitem{Weinberg:1979sa}
  S.~Weinberg,
  Phys.\ Rev.\ Lett.\  {\bf 43}, 1566 (1979).


\bibitem{Pascoli:2003rq}
  S.~Pascoli, S.~T.~Petcov and C.~E.~Yaguna,
  Phys.\ Lett.\ B {\bf 564}, 241 (2003)
  [hep-ph/0301095].


\bibitem{Davidson:2006bd}
  S.~Davidson and F.~Palorini,
  Phys.\ Lett.\ B {\bf 642}, 72 (2006)
  [hep-ph/0607329].


\bibitem{Hambye}
  M.~B.~Gavela, T.~Hambye, D.~Hernandez and P.~Hernandez,
  arXiv:0906.1461 [hep-ph].



\bibitem{MEG}
  M.~Grassi  [MEG Collaboration],
  Nucl.\ Phys.\ Proc.\ Suppl.\  {\bf 149} (2005) 369.


\bibitem{hisano}
  J.~R.~Ellis, J.~Hisano, M.~Raidal and Y.~Shimizu,
  Phys.\ Rev.\ D {\bf 66} (2002) 115013
  [hep-ph/0206110].


\bibitem{profumo}
  A.~Masiero, S.~Profumo, S.~K.~Vempati and C.~E.~Yaguna,
  JHEP {\bf 0403} (2004) 046
  [hep-ph/0401138].

\bibitem{Herrero}
  S.~Antusch, E.~Arganda, M.~J.~Herrero and A.~M.~Teixeira,
  JHEP {\bf 0611} (2006) 090
  [hep-ph/0607263].


\bibitem{Yanagida}
  M.~Fukugita and T.~Yanagida,
  Phys.\ Lett.\ B {\bf 174}, 45 (1986).

\bibitem{Cirigliano:2006nu}
  V.~Cirigliano, G.~Isidori and V.~Porretti,
  Nucl.\ Phys.\ B {\bf 763} (2007), 228 [hep-ph/0608123].

\bibitem{Branco:2006hz}
  G.~C.~Branco, A.~J.~Buras, S.~Jager, S.~Uhlig and A.~Weiler,
  hep-ph/0609067;

\bibitem{Cirigliano:2007hb}
  V.~Cirigliano, A.~De Simone, G.~Isidori, I.~Masina and A.~Riotto,
  JCAP {\bf 0801} (2008) 004
  [arXiv:0711.0778 [hep-ph]].







\bibitem{Grinstein:2006cg}
  B.~Grinstein, V.~Cirigliano, G.~Isidori and M.~B.~Wise,
   Nucl.\ Phys.\ B {\bf 763} (2007) 35 [hep-ph/0608123].


\bibitem{AS}
  N.~Arkani-Hamed and M.~Schmaltz,
  Phys.\ Rev.\  D {\bf 61}, 033005 (2000)
  [arXiv:hep-ph/9903417].

\bibitem{RSoriginal}  
   T.~Gherghetta and A.~Pomarol,
  Nucl.\ Phys.\  B {\bf 586}, 141 (2000)
  [arXiv:hep-ph/0003129].
  S.~J.~Huber and Q.~Shafi,
  Phys.\ Lett.\  B {\bf 498}, 256 (2001)
  [arXiv:hep-ph/0010195];


\bibitem{aps}
  K.~Agashe, G.~Perez and A.~Soni,
  Phys.\ Rev.\  D {\bf 71}, 016002 (2005)
  [arXiv:hep-ph/0408134];
  Phys.\ Rev.\ Lett.\  {\bf 93}, 201804 (2004)
  [arXiv:hep-ph/0406101].



\bibitem{NMFV}
  K.~Agashe, M.~Papucci, G.~Perez and D.~Pirjol,
  arXiv:hep-ph/0509117;

\bibitem{Contino}
  R.~Contino, T.~Kramer, M.~Son and R.~Sundrum,
  JHEP {\bf 0705} (2007) 074
  [arXiv:hep-ph/0612180].


\bibitem{Davidson:2007si}
  S.~Davidson, G.~Isidori and S.~Uhlig,
  Phys.\ Lett.\  B {\bf 663}, 73 (2008)
  [arXiv:0711.3376 [hep-ph]].

\bibitem{Cacciapaglia:2007fw}
  G.~Cacciapaglia, C.~Csaki, J.~Galloway, G.~Marandella, 
  J.~Terning and A.~Weiler,
  JHEP {\bf 0804} (2008) 006
  [arXiv:0709.1714 [hep-ph]].

\bibitem{Csaki:2008eh}
  C.~Csaki, A.~Falkowski and A.~Weiler,
  arXiv:0806.3757 [hep-ph].




\bibitem{Nelson:2000sn}
  A.~E.~Nelson and M.~J.~Strassler,
  JHEP {\bf 0009} (2000) 030
  [arXiv:hep-ph/0006251].


\bibitem{Csaki1}
  C.~Csaki, A.~Falkowski and A.~Weiler,
  JHEP {\bf 0809}, 008 (2008)
  [arXiv:0804.1954 [hep-ph]].

\bibitem{Blanke1}
  M.~Blanke, A.~J.~Buras, B.~Duling, S.~Gori and A.~Weiler,
  JHEP {\bf 0903}, 001 (2009)
  [arXiv:0809.1073 [hep-ph]].

\bibitem{Bauer:2008xb}
  M.~Bauer, S.~Casagrande, L.~Grunder, U.~Haisch and M.~Neubert,
  arXiv:0811.3678 [hep-ph].
 
\bibitem{noi}
  O.~Gedalia, G.~Isidori and G.~Perez,
  arXiv:0905.3264 [hep-ph].



\bibitem{Blanke:2008yr}
  M.~Blanke, A.~J.~Buras, B.~Duling, K.~Gemmler and S.~Gori,
  JHEP {\bf 0903} (2009) 108
  [arXiv:0812.3803 [hep-ph]].


\bibitem{Fitzpatrick:2007sa}
  A.~L.~Fitzpatrick, G.~Perez and L.~Randall,
  arXiv:0710.1869 [hep-ph].

\bibitem{Csaki:2009wc}
  C.~Csaki, G.~Perez, Z.~Surujon, A.~Weiler,
  arXiv:0907.0474 [hep-ph]





\end{thebibliography}
\end{document}